# *Computing the Field*
## *in*
## *Proteins and Channels*

*August 3, 1995*

Bob Eisenberg
Dept. of Molecular Biophysics and Physiology
Rush Medical College
1750 West Harrison
Chicago IL 60612

Predicting function from structure is a goal as old as molecular biology, indeed biology itself: the first glimpse of the anatomy of an animal, like the first glimpse of the anatomy of a protein, must have raised the question "How does it work?" The recent outpouring of protein structures (e.g., Creighton, 1993), often of enzymes in atomic detail (Singh & Thornton, 1995), has raised the same question again and again, and in a most frustrating way, at least for me, because the question is so often answered by description, not analysis. Of course, every biologist knows that a complex structure must be described before its biological function can be analyzed.[1] But every physicist knows that description is not understanding. Indeed, if description is mistaken for understanding, vitalism will soon replace science.

Understanding enzymes is an important goal of modern biology because enzymes catalyze and control most of life's chemistry. Enzymatic action depends on the diffusion of substrate and product, the conformation change of the enzyme, and the quantum chemistry of its active site. All must be analyzed in terms of physical models if enzymatic function is to be understood. The quantum chemistry of systems like enzymes is not well understood because enzymes are flexible structures of many atoms in a condensed phase, containing little empty space, and so are difficult to analyze with Schrödinger's equation (Schatz & Ratner, 1993; Bader, 1990; Parr & Yang, 1989). The movements of the substrate, product, and enzyme must also be understood, presumably as a result of electromechanical models, like Langevin equations (Kramers, 1940; Gardiner, 1985; Hynes, 1985; 1986; Hänggi, Talkner & Borkovec, 1990; Fleming & Hänggi, 1993). Movements of substrate and product have been studied in this tradition, but movement of the protein (i.e., conformation changes) have rarely been connected to the physics that govern them.

I suspect that the idea of conformation change was originally introduced to seek "… the ultimate source of the autonomy, or more precisely, the self-determination that characterizes living beings in their behavior" (Monod, 1972: p. 78[2]). But the description of the conformation change (as allosteric, or whatever) is not the same as a physical understanding. Without a physical model of catalysis and conformation change, as well as electrodiffusion, understanding of enzymatic function is not possible.

A physical model of electrodiffusion is feasible and various forms have been used to analyze transport processes in biology for more than a century. Recently, the nonlinear Poisson-Boltzmann equation ($PB_n$ for short), has been widely used to analyze diffusion in enzymes (Honig & Nicholls, 1995; Davis & McCammon, 1990; Hecht, Honig, Shin & Hubbell, 1995, is particularly important because it provides direct experimental verification of the theory; also: Warwicker & Watson, 1982; Klapper, Hagstrom, Fine, Sharp & Honig, 1986; Gilson, Sharp & Honig, 1988; Davis, Madura, Luty & McCammon, 1991). $PB_n$ is a major contribution to our understanding, but $PB_n$ is not a theory of enzymatic function, because it describes only electrodiffusion, not catalysis or conformation change. The question then arises whether a significant biological function like transport can be understood *just* from an understanding of electrodiffusion?

Understanding membrane transport has been an important goal of biology for more than a century because it governs so much of life. In particular, membranes—in their role as gatekeepers to cells—are responsible for signaling in the nervous system; for co-ordination of the contraction of skeletal muscle and the heart, allowing its muscle to function as a pump. Membranes contain receptors or effectors for many drugs and natural substances that control the life of cells.

---

[1] Linnaeus must precede Darwin; or rather "Voyage of the Beagle" must precede "Origin of Species".
[2] in a chapter entitled 'Microscopic Cybernetics'.

Viewing biological transport as a form of diffusion was, however, only partially successful (Hille, 1992; Hille, 1989). Analysis usually assumed that diffusion occurred in systems of fixed structure, in particular of fixed cross sectional area, but we now know that the area for diffusion is modulated and controlled in most biological systems by the opening and closing of pores, namely channels in membrane proteins. Indeed, determining the number of open channels and their modulation is a main task of physiologists nowadays (Alberts *et al.*, 1994; Hille, 1992). Of course, the conformation change that opens a channel itself needs to be understood in physical terms with explicit electromechanical models like Langevin equations (describing the motion of the atoms of the protein, channel, and permeating molecule), but once a channel is open, the role of gating is much reduced.

**THE OPEN CHANNEL**. An open channel forms a well defined structure of substantial biological importance, whose function should be "wholly interpretable in terms of specific chemical [or physical] interactions…" (Monod, 1972, p. 78). The age old question "How does it work?" should be easier to answer when 'it' is an open channel, than when it is anything else, at least of such general biological importance.

The starting place[3] for a theory of open channels is a theory of electrodiffusion rather like that used previously to describe membranes. The theory uses Poisson's equation to describe how charge on ions and the channel protein creates electrical potential; it uses the Nernst-Planck equations to describe migration and diffusion of ions in gradients of concentration and electrical potential. Combined[4], these are also the "drift-diffusion equations" of solid state physics, which are widely, if not universally used to describe the flow of current and the behavior of semiconductors (Ashcroft & Mermin, 1976) and solid state devices, like transistors (Sze, 1981, Selberherr, 1984). The drift-diffusion equations describe the shielding or screening of permanent or fixed charge whereby the ions in the ionic atmosphere in and around a (channel) protein help determine the potential profile of its pore, a phenomena long known to be biologically important (Frankenhaeuser & Hodgkin, 1957; McLaughlin, 1989; Green & Andersen, 1991). Mathematical difficulties have been limiting, however, and so attention has usually been focused on the ionic atmosphere at the surface of the membrane or ends of the channel, and not the co– and counter ions within the channel's pore.

Many theories focus on systems at equilibrium in which all fluxes are essentially zero. In the latter case, the PNP equations reduce to the (one dimensional) $PB_n$ equations. A great deal of important work has been done on $PB_n$, and in some ways $PB_n$ is the most physical theory of proteins now being used; nonetheless, it is of limited use in understanding the open channel because the natural biological function of channels occurs only away from equilibrium (as does the biological function of most enzymes!). The biological function of both channels and enzymes is usually flux.

Significant flux flows even at the reversal potential of a typical, imperfectly selective channel.

---

[3] Electrodiffusion in biology occurs in a liquid in which friction and interactions predominate. As Berry, Rice & Ross put it in their textbook (1980, p. 844, emphasis added): "…the principal difference between a dilute gas and a liquid …[ is that] in a dilute gas a typical molecule is usually outside the force fields of all other molecules and only occasionally in the force field of one other molecule (binary collision), whereas in a liquid a typical molecule is usually within the force field of, say, 10 nearest-neighbor molecules *and is never completely free of the influence of other molecules*." The starting place for a theory of channels must be a theory of liquids. Theories based on gas phase kinetics, like Eyring rate theory, are likely to miss the important effects of interactions (i.e., friction), which dominate the properties of liquids, but not gases.

[4] We call the combination PNP to emphasize the importance of the electric potential and the analogy with solid state physics.

Fluxes are all zero only in a perfectly selective channel at its reversal potential, which is then indeed an equilibrium potential. A generalization of the PB$_n$ equations is needed to predict flux and PNP is one such generalization. Interestingly, channels at equilibrium have a quite restricted repertoire of behavior, as do semiconductors (which cannot be transistors in the absence of flux), and so an equilibrium theory can give only limited insight into the repertoire of natural functions of channels, even if it could calculate one.

Many theories of ion movement include only some types of charge (e.g., permanent charge is usually ignored in the electrochemical literature cited below). But all charge is likely to have global effects, at least judging from work on charge at the ends of channels (Green & Andersen, 1991). Thus, the whole system, containing all types of charge and flux, has to be analyzed if the biological function of the open channel is to be predicted successfully from its structure.

***PNP THEORY TODAY***. The mathematical difficulties in the analysis of the full system have been largely overcome. PNP theory can now predict the current through an open channel given its structure and distribution of fixed (i.e., permanent) charge. Indeed, once the structure of the open channel is known, and thus the distribution of its permanent charge, along with diffusion constants, PNP theory predicts its properties—the fluxes and current through it—in all experimental conditions of varying concentrations and trans-membrane potentials.

Preliminary work shows that PNP theory fits a wide range of data, taken from many solutions, that is difficult to fit with traditional models (Chen *et al*., 1995*a,b*; Franciolini & Nonner, 1994*a,b*; Kienker & Lear, 1995; Kienker, DeGrado & Lear, 1994). PNP automatically predicts a wide repertoire of behavior, because it is nonlinear and the potential profile in the channel and its pore (which is an output, not assumption, of the theory) changes significantly with experimental conditions. A qualitative understanding of this behavior is possible in many cases, but PNP is a mathematical theory, a set of coupled nonlinear differential equations, describing interactions arising from all types of charge, and flux. It is not always possible to rationalize the behavior of such systems in a few words, particularly if several terms, or types of charge (some positive, some negative) are significant.[5]

PNP theory fits a wide range of data because shielding usually has global effects, spreading across the entire channel. The potential and contents of the channel's pore (i.e., the ionic atmosphere of counter and co–ions within the channel) change as solutions or membrane potential are changed, as they must, if potential and concentrations simultaneously satisfy Poisson and Nernst-Planck equations. Our calculations show changes of potential of several $k_BT/e$ in most locations when typical solution changes are made. In loose terms, we can say that small changes in net charge in and near the channel make significant changes in potential and even bigger changes in flux: potential is a sensitive function of net charge, and flux depends exponentially on potential[6]. The ionic atmosphere and shielding are major determinants of a channel's properties.

---

[5] Before trying to summarize the physics and properties of PNP, one should count the number of pages used by textbooks to describe and rationalize the behavior of a transistor, even as it works in a ***restricted range of bias voltages and currents*** (Ashcroft & Mermin, 1976; Sze, 1981; Selberherr, 1984). Of course, if the bias voltages are changed, the qualitative properties of the transistor change, for example, from an amplifier, to a switch or limiter, and each qualitative property requires extensive discussion, because each property arises from a different distribution of electric field and flux of holes and electrons, even though the profile of permanent charge remains the same.

[6] Eq. (2) illustrates the dependence of flux on potential and concentration within the channel. Eq. (9) illustrates the dependence of concentration on potential. Understanding the dependence of potential requires numerical solution of the full set of PNP equations, see Appendix, p. 26 and footnote 11.

PNP theory is significant as a theory of an important biological phenomena—open channel permeation—arising from a simple physical mechanism, electrodiffusion. But it is also important in the more general context of proteins and enzymes as well, because PNP theory shows by implication, if not derivation or proof, that any property of a protein will be strongly influenced by changes in the electric field, whether gating of a channel, mediated or active transport of a 'permease', conformation changes in an enzyme, or catalysis itself. Indeed, we suspect that many of these processes will be dominated by the electric field and its change in shape. PNP adds another example[7] to those already known, in which the electric field dominates the biological function of a protein (Honig & Nicholls, 1995; Davis & McCammon, 1990; see also Warshel, 1981; Warshel & Russell, 1984; Warshel & Åqvist, 1991).

***ELECTRODIFFUSION IN SEMICONDUCTORS, SOLUTIONS, AND CHANNELS.*** The fundamental physical process in transistors and semiconductors is the migration and diffusion of charged quasi-particles—holes and electrons—in electric fields and gradients of concentration, just as the fundamental process in channels is the diffusion of ions, and perhaps quasi-particles, as well. In electrochemistry and semiconductor physics, the electric field is usually described by Poisson's equation[8] (that specifies how charge creates potential)

$$-\varepsilon_{H_2O}\varepsilon_0 \frac{d^2\varphi}{dx^2} = \overbrace{eP(x)}^{\text{Permanent Charge}} + \overbrace{e\sum_j z_j C_j(x)}^{\text{Channel Contents}} + \underbrace{\overbrace{\tilde{\varepsilon}\left[\Delta(1-x/d) - \varphi(x)\right]}^{\text{Induced Charge}}}_{\text{Deviation from Constant Field}} \tag{1}$$

Electrodiffusion is usually described by the Langevin equation (that describes individual trajectories: Kramers, 1940; Gardiner, 1985; Eisenberg, Klosek & Schuss, 1995) or the Nernst-Planck equations (that describe the probability density function of these trajectories). Here, the Nernst-Planck equations are written in the integrated form we have found most useful.

$$J_j = D_j \frac{C_j(L)\exp(z_j V_{appl})}{\int_0^d \exp z_j \Phi(\zeta) d\zeta} - D_j \frac{C_j(R)}{\int_0^d \exp z_j \Phi(\zeta) d\zeta}; \qquad I = \pi r^2 \cdot \sum_j e z_j J_j \tag{2}$$

The Nernst-Planck equations are fundamentally nonlinear because the conductance of ionic solutions depends on concentration.[9] Thus, as the concentration changes, the migration of ions changes, even if everything else is constant. The Poisson equation is nonlinear (in the present context) because the net charge of ions is one of the source terms in the equation, but that term depends on the potential, the output of Poisson's equation. The Nernst-Planck equations are also nonlinear because they depend on the potential profile $\dot{\,}$ which in turn depends on everything else in the system, through the Poisson equation.

---

[7] a particularly convincing and biologically relevant example, we immodestly think.
[8] variables, dimensions of variables, brief explanation and references can be found in the Appendix.
[9] This essential point is easily obscured by the several forms of equations and the plethora of parameters used to describe migration—believe it or not, conductance, conductivity, mobility (conventional and absolute), and diffusion constant are all used in *elementary* texts (Bockris & Reddy, 1970; Bard & Faulkner, 1980) and advanced treatments (e.g., Justice, 1983)! The essential point is that the conductance of a given volume of solution depends on the concentration of ions in the solution and the conductance appears as a coefficient of a derivative in the Nernst-Planck differential equations. Concentration varies over a wide range in most systems and so linearized approximations cannot describe their qualitative properties.

The nonlinearity of the Poisson and Nernst-Planck equations allows a richness of behavior that we use every day, given the role of solid-state electronics in our technology and economy. Transistors are semiconductors designed to do specific jobs, to have nonlinear properties that arise in part from the nonlinearity of the Nernst-Planck equations, in part from the nonlinearity of their coupling to the Poisson equation discussed later in this paper. The distribution of permanent charge is chosen by the designer of the transistor to create the shape of the electric field he wishes, thereby making an otherwise uninteresting homogeneous lump of pure silicon into a switch, an amplifier, a detector. An entire computer can be built solely out of transistors that obey the Nernst-Planck (and Poisson) equations in two (or three) dimensions. All written knowledge, and all mathematical operations, can be stored or executed by a computer, and so mathematical solutions to the Nernst-Planck (and Poisson) equations can have a rich range of behavior!

The PNP equations can themselves produce the microscopic cybernetics thought by some philosophers of science to be characteristic (or defining) of life (Monod, 1972, p. 68-80), even in the absence of the allosteric conformation changes they had postulated. Of course, no one yet knows if channels actually perform these cybernetic functions by changing the shape of the electric field, and if such functions are important for the life of the animal as a whole, as likely as it seems, given their importance in semiconductor technology.

***TRADITIONAL MODELS OF OPEN CHANNELS: RATE CONSTANTS.*** Traditional models usually use rate constants to describe open channel permeation, or gating (Andersen & Koeppe, 1992; Hille, 1992) or changes in conformation in general (Hill, 1977; Walsh, 1979; Hill, 1985). Traditional models usually describe the structure of channel proteins as a distribution of potential, 'a potential of mean force' which in turn determines its rate constants.

It is hard to imagine a theory in which a rate constant is independent of the potential profile. Indeed, I am unaware of a rate theory in which rate constants are independent of the underlying potential of mean force. Even in the simplest most approximate theories of gas phase kinetics—like Eyring rate theory—rate constants are exponential functions of barrier height (Wigner, 1938; Laidler, 1969; Johnson, Eyring & Stover, 1974; Truhlar & Garrett, 1984; Truhlar, Isaacson & Garrett, 1985; Skinner & Wolynes, 1980). In more general theories—which provide the basis for rate theories in condensed phases and from which they must be derived (Hynes, 1985; 1986; Berne, Borkovec & Straub, 1988; Fleming & Hänggi, 1993)—the rate constant also depends *exponentially* on potential (*cf.* Eisenberg, Klosek & Schuss, 1995, and eq. (11) & (12) of the Appendix). In either case, anything that changes the potential profile will change the rate constant, often dramatically. Thus, rate constants of traditional theories—whether of gating, permeation, or 'active' transport—are likely to vary with experimental conditions, just as potential profiles vary.[10]

A theory that uses rate constants independent of the potential profile (or of concentration and *trans*-membrane potential) will not be able to describe behavior produced by changes in the field and thus may fail to predict many important channel phenomena.

***SEMICONDUCTORS, SOLUTIONS, AND CHANNELS.*** The analogy between semiconductors and ionic solutions has been known for some time but it has not been very productive because semiconductors contain permanent charge (doping) and ionic solutions do not. A (particular) distribution of permanent charge is what turns a semiconductor into a transistor, for example, and ionic solutions cannot be transistors because they do not have permanent charge[11].

---

[10] If the permeation pathway is occluded in a closed channel, one would expect permeant ions to have much less effect on the opening process than on open channel permeation, or on the closing process, for that matter.

Significant charge is more widely distributed and concentrated in proteins than is sometimes realized (see p. 20). Permanent charge is found in most atoms and bonds of proteins, not just in atoms with formal charges; many atoms of proteins contain between 0.1 and 0.6 elementary charges, for example, the carbon, nitrogen, oxygen, and perhaps even hydrogen in the amide bonds that link every amino-acid in a protein (Schultz & Schirmer, 1979; Fersht, 1985; McCammon & Harvey, 1987; Brooks, Karplus & Pettitt, 1988, Creighton, 1993). The charge on each of these atoms is nearly as significant as a formal charge on a carbonyl or amine group of the protein, or, for that matter, on a permeating ion like $Na^+$ or $Cl^-$.

**SOURCES OF DIFFUSION.** Diffusion is driven by concentrations which are the sources of mass and free energy, if we use language of thermodynamics; they are the sources of flux, if we use the language of $19^{th}$ century physics; they are the source of trajectories, if we use the language of probability theory and stochastic processes.

In biological membranes and channels, the concentration gradients arise from ions in the baths adjoining the membrane. The concentrations of these ions are maintained by ancillary experimental or biological systems that supply the ions equivalent to those that move through the channel and so sustain the free energy of the baths.

The concentrations in the baths are, in fact, the only sources for diffusion; ions are not supplied within the channel, nor can they appear there spontaneously. Semiconductors are a little different because of the recombination process, but this is usually ignored in theories. Recombination does not occur in channels or solutions containing only strong electrolytes like $Na^+$, $K^+$, $Ca^{++}$, or $Cl^-$. It can occur in other situations, see p. 10.

**SOURCES OF THE ELECTRIC FIELD.** The sources for the electric field that drive the drift (i.e., migration) of ions are more complex than the sources for diffusion, whether in semiconductors, solutions, or channels. The sources are the several kinds of electrical charge in the system, each type with its own properties.

For example, only the charge in the baths, on the boundary of the system, connected to amplifiers, pulse generators or batteries, maintains the *trans*-membrane potential. That charge must be maintained by a continuous supply of energy from the outside world because flux flows across the channel dissipating energy and producing heat. The other types of charge (described below) also help create the electric field. They, however, cannot supply energy in the steady-state because they are not on a boundary of the system and so are not connected to an energy source outside the system.

In free solution, the dominant charge is usually the mobile charge (i.e., ions) supplied at the boundaries of the system (i.e., at the electrodes) by the experimental apparatus that maintains the electrode potentials.

In channels or semiconductors, the dominant charge is usually the permanent charge because it has such a high density, many molar if described as a volume charge density, even if formal charges are absent (see p. 6 and 20). The qualitative properties of the electric field in channels (and semiconductors), as well as its quantitative current voltage (IV) relations, are determined by the distribution of permanent charge, with a strong assist from the mobile charge (the ions) sup-porting the *trans*-membrane potential (analogous to the bias potential of a two terminal transistor).

---

[11] When $P(x) = 0$ and $\tilde{\varepsilon} = 0$, the PNP equations describe current flow in free solution (Bockris & Reddy, 1970). Syganow & von Kitzing, 1995, and Park *et al.,* 1996, describe some of the conditions under which ohmic behavior can arise from these nonlinear equations.

Charge create the electric field according to Coulomb's law (if the charge is discrete) or Poisson's equation (if the charge is distributed). The equations of the classical electric field describe the mapping between charge and potential both in the macroscopic world and in quantum chemistry (Hellman-Feynmann theorem: Feynmann, 1939; Deb, 1981, Mehra, 1994, p. 71; *cf.* Bader, 1990; Parr & Yang, 1989). *All* charge must be included in whichever equation(s) apply, because the effects of the dominant charge are significantly modified by the other charges. Linking potential and charge requires simultaneous solution of Poisson's equation and the Nernst-Planck equations, subject to boundary conditions: these equations automatically describe the interactions of the different types of charge and the fields they create.

**POISSON'S EQUATION IN SEMICONDUCTORS.** Poisson's equation has been used for some 40 years by solid-state physicists to describe the electrical field in semiconductors (Roosbroeck, 1950, and Shockley, 1950). Coupled to the Nernst-Planck equations, Poisson's equation describes the many semiconductor devices with a wide range of characteristics. These equations are coupled to each other because the ions that create the electric field (as described by Poisson's equation) flow and so are modified by the electric field (as described by the Nernst-Planck equations).

The combined Poisson and the Nernst-Planck equations have been derived and are used in several different experimental and theoretical traditions (Ashcroft & Mermin, 1976; Mason & McDaniel, 1988; Spohn, 1991; Balian, 1992; Jerome, 1995; at equilibrium see references to $PB_n$, most importantly, Hecht, Honig, Shin & Hubbell, 1995; also, Honig & Nicholls, 1995; Davis & McCammon, 1990), where they have been explicitly tested many times by simulation and experiment. The drift diffusion equations are tested implicitly every time we use a semiconductor device designed with them. There is little question in these fields that the equations are consistent with the laws of mathematics and physics, and (in particular) that they handle self-energy correctly[12]; for example, energetics are critically analyzed *and compared to experiments* at length in the literature of $PB_n$ (e.g., MacKerell, Sommer & Karplus, 1995; Antosiewicz, McCammon & Gilson, 1994; Gilson *et al.*, 1993; Sharp & Honig, 1990; Jayaram *et al.*, 1989; Gilson & Honig, 1988).

We will discuss the validity of PNP as a description of open ionic channels in some detail later, starting on p. 15. Suffice it to say here that the PNP equations have proven helpful in many fields, even where they are only approximately valid and the microscopic meaning of their parameters are not fully understood. The theory has been widely used even though it consists of macroscopic continuum equations that describe charge as a fluid, not as the swarm of discrete particles and quasi-particles (e.g., holes and electrons) it really is.

**ELECTRICAL FORCES IN CHEMISTRY.** Electric fields are so important for channels because electrical forces dominate chemical phenomena. Indeed, in a certain sense, all of chemistry arises from electric charge and its interactions according to Coulomb's law:

" ... all forces on atomic nuclei[13] in a molecule can be considered as purely classical attractions involving Coulomb's law. The electron cloud dis-

---

[12] Jakobsson, 1993, p. 34, feels otherwise, however.

[13] Nuclei contain nearly all the mass of atoms, while occupying a negligible volume, and electrons move where the nuclei move and carry them (in the Born-Oppenheimer approximation: Parr & Yang, 1989). Thus, forces on nuclei determine the acceleration of atoms, both of their nuclei and electrons.

tribution is prevented from collapsing by obeying Schrödinger's equation."
(Feynmann, 1939).

For this reason, any description of a protein must describe its electron cloud and nuclei, which form the distribution of electrical charge in the protein: its permanent charge. Any description of the movement of ions through a channel protein should keep track of all the charges in and around the channel for the same reason: the electrical charge determines the forces on the permeating ions, as well as the forces that make the protein move, the forces that drive its conformational changes.

**CHARGE IN CHANNELS.** The electric field in a channel arises from charge. PNP theory includes four types of charge,

(1) the permanent charge on the channel protein that arises from its chemical structure. Permanent charge is often called bound charge (Griffiths, 1981; Purcell, 1985, Ch. 10) but that phrase is unfortunate because mobile charge is also bound to proteins, in a very real sense. Permanent charge is sometimes called fixed charge, but that phrase is also unfortunate (at least on the atomic scale) because the permanent charge is highly flexible, see p. 16.

(2) the contents of the channel, namely, the charge in the channel's pore carried by the (average) concentration of ions there. The concentration of ions in the channel's pore both determines and is determined by the electric field.

The double role of charge, as cause and effect of the field, can easily cause difficulties. That is exactly why the electric field was so difficult for $19^{th}$ century physicists to understand, according to several historical accounts (Buchwald, 1985; Hunt, 1991; Siegel, 1991). Indeed, until the discovery of the electron, at the end of that century, the distinctions between permanent and induced charge were not understood. The electric field is, however, no more complicated than gravitational interactions of deformable (*cf.* 'polarizable') objects of similar mass, like binary stars. There, too, the location and shape of the sources is both determined by and determines the gravitational force.

Gravitation is not usually considered in this light (except by astronomers) because we usually compute gravitational forces between rigid objects of very different mass. Then, only one object determines the field, to a good approximation. The essential fact about electricity is that this approximation rarely applies: most objects, e.g., atoms and ions, have about the same amount of charge and many can move and deform ('polarize') in the electric field.)

(3) the dielectric charge (that we call induced or polarization charge: Griffiths, 1989, Ch. 4; Purcell, 1985, Ch. 10) which is traditionally described by the volume density of (hypothetical) dipoles (of infinitesimal size) that represent the small movements of electrons and nuclei (etc.) induced by the electric field. Dielectric or polarization is defined as the induced charge movement (dipole moment per unit volume, to be precise) proportional to the local electric field.

Induced charge with more complex properties is usually described as a component of mobile charge that varies with the electric field. The ions inside a channel's pore form a concentration of mobile charge that varies nonlinearly with the electric field.

Polarization or induced charge does *not* include dipoles independent of the electric field, so it does not include the macroscopic dipoles of say a carbonyl group, even though carbonyls are invariably called a 'polar' group! In our system of definitions, that kind of charge is described as part of the permanent charge.

(4) boundary charge, namely, charge applied at the boundaries of the system during the experiment (by electronic and chemical apparatus) to keep the potential and concentrations (nearly) constant at those places.

Everyone who does experiments knows how important it is to measure and/or control the concentrations and potentials on both sides of channels that together determine the thermodynamic driving force, the gradient of electrochemical potential. These concentration and potentials supply the energy, charge, and matter that make currents and fluxes flow. They are the sources of free energy for channel phenomena and so, if they vary, nearly everything measured also varies.

For precisely the same reasons, theory must describe these sources carefully. In the present case (and usually) the sources are on the boundaries of the system, and are given by boundary conditions, when the system is described in three dimensions.

**SHIELDING OF 'SURFACE' CHARGE.** Permanent charge on the surface of membranes (or at the ends of channels) attracts ions of opposite sign from the (overall) electrically neutral bathing solution, and thereby creates an ionic atmosphere with net charge (called the diffuse double layer or Gouy-Chapman layer, see McLaughlin, 1989, and Green & Andersen, 1991, for extensive references in the channel context; Bockris & Reddy, 1970; Bard & Faulkner, 1980, are gateways to the electrochemistry literature) analogous to the ionic atmosphere of Debye-Hückel theory. This charge produces a potential drop in the surrounding bathing solution, called the surface potential in physical chemistry, and the built-in potential in semiconductor physics. If the concentration of ions in the bath is very high, compared to the density of surface charge, the ionic atmosphere 'shields' or 'screens' the surface charge, and the surface or Donnan potential extends only a short distance into the bath. If the concentration of ions is low, the surface charge is hardly shielded, and the Donnan potential extends a far into the solution.

In semiconductors, screening has long been known to be important. That literature of thousands of papers can be reached through Selberherr, 1984; Rubinstein, 1990, and Jerome, 1995. In electrochemistry, screening has not been analyzed as carefully, usually because models have been reduced to avoid mathematical difficulties by ignoring one of the types of charge or another (*cf.* p. 8), or by setting flux to zero (e.g., Mafé, Pellicer & Aguilella, 1986, 1988; Mafé, Manzanares & Pellicer, 1988, 1990; Murphy, Manzanares, Mafé & Reiss, 1992; Guiraro, Mafé, Manzanares & Ibáñez, 1995; Brumleve & Buck, 1978; Nahir & Buck, 1993; de Levie & Moreira, 1972, de Levie, Seidah & Moreira, 1972; de Levie & Seidah, 1974; de Levie, Seidah & Moreira, 1974, who used the name Poisson-Nernst-Planck).

In channology, the effect of surface charge on open channel permeation has also received much attention, see Apell, Bamberg, Alpes & Läuger, 1977; McLaughlin, Mulrine, Gresalfi, Vaio & McLaughlin, 1981; McLaughlin, Eng, Vaio, Wilson & McLaughlin, 1983; Dani, 1986; Green, Weiss & Andersen, 1987; Jordan, 1987; Kell & DeFelice 1988; Peskoff & Bers, 1988; Jordan *et al.*, 1989; McLaughlin, 1989; Cai & Jordan, 1990; Mathias, Baldo, Manivannan & McLaughlin, 1991; Green & Andersen, 1991.

In channology, the effect of surface charge on gating phenomena has been studied for many years (at least since Frankenhaeuser & Hodgkin, 1957; Chandler, Hodgkin & Meves, 1965; Gilbert & Ehrenstein, 1984, is a review; see also Hille, 1992). Divalent ions and pH also have large effects on gating. Both have been explained by their effect on the surface potential (McLaughlin, Mulrine, Gresalfi, Vaio & McLaughlin, 1981*;* McLaughlin, 1989; Green and Andersen, 1991).

In channology, the effect of shielding of the protein's charge in the channel's pore has received less attention, although some reduced models have been developed (e.g., Teorell, 1953;

see also Mauro, 1962 Bruner, 1965*a*, 1965*b*, 1967; Åqvist & Warshel, 1989; Edmonds, 1994) and Peng, Blachly-Dyson, Forte & Colombini, 1992, and Zambrowicz & Colombini, 1993, apply Teorell's theory (in essence) to modern data.

In channology (or anywhere else), what has been very difficult to develop has been a computable treatment of shielding that includes flux and allows coupling of the effects of all types of charge.

Only recently have computers been fast enough, and numerical techniques sophisticated enough to analyze the entire system, involving as it does coupled nonlinear partial differential equations describing flux in the presence of surface, induced, and fixed (i.e., permanent) charge (first successfully treated, in semiconductors, Fatemi, Jerome & Osher, 1991; later, independently, in channels, by Chen, *cf.* Chen & Eisenberg, 1993*b*). Our preliminary results using these new methods reinforce what others have suggested (Green and Andersen, 1991): surface charge has global effects; all types of charge in the channel interact with each other. Thus, understanding and predicting the effects of bath concentration and *trans*-membrane potential requires the simultaneous solution of a system of nonlinear differential equations (Poisson and Nernst Planck equations) and boundary conditions, including all types of charge. A reduced model seems impossible[14]: subtraction of surface potential and constant field (i.e., induced charge) terms from the potential leaves a residue due to the permanent charge, that would be independent of concentration and *trans*-membrane potential in a linear system. Here, however, in the nonlinear world of PNP, calculations show that the residue depends nearly as strongly on experimental conditions as the total potential (Chen and Eisenberg, personal communication).

***CHANGES IN PERMANENT CHARGE.*** In most situations the distribution of permanent charge on a protein does not change. The exceptions, however, are of considerable biological importance. When the conformation of the channel protein changes for whatever reason, its distribution of permanent charge will change. When transmitters or second messengers bind to a protein (or phosphorylate it), they add or subtract charge and are likely to induce a conformational change. Both will change the distribution of permanent charge significantly and thereby change the electric field even if they are formally neutral, because nearly all transmitters and messengers are zwitterions or polar (otherwise, they would not be very soluble in water!). Indeed, a conformation change in any protein near a channel will almost certainly modulate the electric field in the nearby lipid membrane and adjacent channels, thereby initiating and governing gating and open channel permeation.

When the chemical nature of the protein changes because of changes in covalent bonds, the distribution of permanent charge will also change. Obvious examples are phosphorylation (that produces local concentrations of $P_i = \left[H_2PO_4^-\right] + \left[HPO_4^-\right]$) and changes in ionization of the acidic or basic groups in the protein (that produce local concentrations of $H_3O^+$ or $OH^-$) caused by changes in pH. If phosphorylation (or ionization, see next paragraph) actually occurs within a channel, they will change the concentration and species of current carriers there, as well as permanent charge. In that case, the dragging of ions by $P_i$, $H_3O^+$, or $OH^-$ (and *vice versa*) might play an important role in active transport or some types of gating.

***FIELD DEPENDENT IONIZATION.*** Experiments show (Edsall & Wyman, 1958; Tanford & Kirkwood, 1957) that the electric field can significantly change the ionization (i.e., permanent charge) of proteins by changing the effective $pK_a$ of hydrolyzable (often called 'ionizable') groups—Glu, Asp, Arg, Lys, His. The effects found experimentally have been extensively analyzed by the mean field Poisson-Boltzmann ($PB_n$) theory (MacKerell, Sommer & Karplus, 1995; Potter, Gilson, and McCammon, 1994; Antosiewicz, McCammon & Gilson, 1994; Rajasekaran, Jayaram & Honig, 1994; Yang *et al.,* 1992; Straatsma and McCammon, 1991). Of course, *all* the energy terms in $PB_n$ or PNP contribute to the change in $pK_a$. In channel proteins, these include the energy of the membrane potential (i.e., of the charges on the boundary the membrane and in the bath), of the ions in the channel, of the permanent charge, of the induced charge, and of mechanical and electrical forces in the boundary conditions themselves (e.g., the 'dielectric pressure').

---

[14] although analytical approximations to the solutions of PNP can be quite useful (Syganow & von Kitzing, 1995; Park *et al.,* 1996).

The possibility that permanent charge in a channel's pore is modulated by membrane potential or phosphorylation (etc.) is tantalizing because such modulation might provide a link between the elaborate biochemical mechanisms that control channel function and the physics of the open channel (Gilbert & Ehrenstein, 1970; Cramer *et al.,* 1995, p. 628). Ionizable residues are found in putative pore forming regions of many channels (review: Perachia, 1994) and mutations in them often have profound effects on channel function and modulation, both on gating and on open channel permeation (Miller, 1989; Montal, 1995). If significant ionizable charge is present within a channel's pore, complex interactions will occur: charge creates the field, the field modifies the charge, and the potential barriers to permeation will surely change in an interesting way, at least in PNP theory, and probably in the real channel as well, both transiently and at steady-state[15]. In my view, the modulation of the structure of the potential barriers has a good chance of being part of the mechanism of voltage dependent gating and a reasonable chance of being part of the mechanism of other types of gating as well.

**ENERGY IN ELECTRODIFFUSION SYSTEMS.** The definition and computation of energy must be correct if it is to be used to predict the ionization, rate constants, state probabilities, or other properties of proteins or channels.

The idea of energy is considerably more subtle than it sometimes seems, if systems are dissipative, nonlinear, and open, thus violating, in many respects, the assumptions made in elementary treatments of thermodynamics or mechanics. Consideration of the mechanics of nonlinear mechanical systems containing just a few components, like a driven dissipative pendulum (Ben-Jacob, Bergman, Matkowsky & Schuss, 1982; Kupferman, Kaiser, Schuss & Ben-Jacob, 1992), shows the need for careful thought and mathematical treatment before extending the laws of thermodynamics to nonequilibrium nonlinear systems.

Usual treatments of energy in mechanical systems require[16] the boundary conditions of the system to be 'holonomic'[17]. These treatments can be extended to frictional systems if Rayleigh's dissipation function can be defined (Goldstein, 1980, p. 21-25, p. 62). Most treatments of irreversible thermodynamics depend on the definition of a such a function (e.g., Katchalsky & Curran, 1965). Treatments of energy in an equilibrium, quasi-thermodynamic theory like $PB_n$ (*loc. cit.*, and also Antosiewicz, McCammon & Gilson, 1994; Gilson *et al.*, 1993; Sharp & Honig, 1990; Jayaram *et al.*, 1989; Gilson & Honig, 1988) cannot themselves define Rayleigh's function because they assume equilibrium, and at equilibrium, atoms and charges have zero (mean) velocity, velocity dependent potentials do not appear (Goldstein, *loc. cit.*), no flux flows, and so frictional forces and dissipation are zero, as well.

---

[15] It should be mentioned that 'activation curves' showing the voltage dependence of many rate constants of (macroscopic) gating (Hodgkin & Huxley, 1952; Hille, 1992) have the shape of titration curves.

[16] As Goldstein puts it in the second edition of his classic text (p. 64, *op. cit.*): "A good deal has been written about Hamilton's principle for nonholonomic systems, and most of it is wrong (including some things that were said in the first edition). …[Rund, 1966, concludes that] Hamilton's principle is applicable only to holonomic systems."

[17] which for present purposes, can be crudely translated as 'isolated' (Goldstein, 1980, p. 11, also see items listed under 'constraints' in his index).

It is not clear *a priori* whether Rayleigh's dissipation function can be defined for a general electrodiffusion system because diffusion is strongly coupled to the electric field, and the system (with coupling) is inherently nonlinear and open, with flux of energy and matter (and charge) across its boundaries, as we have mentioned many times. The work needed to move charge in a protein or channel (e.g., to ionize residues in a channel) may, or may not be describable by a path independent energy function (even in principle: Griffiths, 1989, p. 187). It remains to be seen.

Whether or not energy and dissipation functions can be defined for these systems—and I surely hope they can!—any theory using energy must keep careful track of all its components:

(1) energy of all the types of charge in the channel (*cf.* p. 8);
(2) chemical energy, i.e., the free energy associated with the concentration, and entropy of each mobile species;
(3) energy losses to friction everywhere in the system and in the boundary conditions, perhaps using a model extended, like the hydrodynamic model, to include heat, energy, and mass flow (Chen, Eisenberg, Jerome & Shu, 1995; Jerome, 1995; Huang, 1987, p. 96);
(4) energy flows in the boundary conditions and the apparatus that maintains them;
(5) energy changes associated with volume changes in the system (Conti, Inoue, Kukita & Stühmer, 1984; Zimmerberg & Parsegian, 1986; Zimmerberg, Bezanilla, & Parsegian, 1990; Peng, Blacly-Dyson, Forte & Colombini, 1992, Vodyanoy, Bezrukov & Parsegian, 1993).

Each component of energy arises in a different part of the system, with its own characteristics. It is not clear which of the components of energy should be included when calculations are made of the rate constants or state probabilities of a channel (or protein), using the 'Boltzmann equation' (Hille, 1992, p. 12). Do the state probabilities of a conformation change (for example) depend only on the electrical energy of the charge that moves during that conformation change in an unvarying profile of potential, as is usually assumed in channology (*loc. cit.*)? Or do the probabilities also depend on the energies of other charges, on the energy lost to friction, on the energy involved in changing the volume or potential profile, and on the energy flow in the boundary conditions?

Difficulties of this sort arise in any quasi-thermodynamic treatment of nonequlibrium systems—not just in $PB_n$ theory—and have motivated me, following in the footsteps of many others (*cf.* footnote 27), to abandon equilibrium models of channels, whether of the equilibrium $PB_n$ or irreversible flavor, and to use kinetic models instead, with the flavor of explicit dynamics, whether molecular, Langevin, PNP, or hydro- dynamics (Elber *et al.,* 1995; Eisenberg, Klosek & Schuss, 1995; Chen & Eisenberg, 1993*a*; Chen, Eisenberg, Jerome & Shu, 1995).

**ALL CHARGE PRODUCES THE ELECTRIC FIELD.** The electric field, that determines the energy, arises from charge according to Poisson's equation, which itself is just a restatement of Coulomb's law.

To describe a channel or protein, either Poisson's equation or Coulomb's law needs to include as its sources all the types of charge described on p.8 because each helps produce the field. The equations need to describe how the charge and potential at the boundaries of the baths and the boundaries of the channel (at its walls) change as experimental conditions change, e.g., as concentrations or potentials are changed in the bath: the sources (boundary conditions) in the theory must, of course, have the same properties as the charge in the physical world.

Interestingly, traditional theories of ionic solutions (see references in Bockris & Reddy, 1970; Bard & Faulkner, 1980) and liquids (see references in Allen & Tildesley, 1990), and simulations of molecular dynamics, even of channels, (reviewed in Roux & Karplus, 1994) often pay little attention to charge on the boundaries[18] and the resulting macroscopic electric fields, and so cannot deal with many phenomena of channels, membranes, cells (or tissues, for that matter), which are produced by charge at the boundary of the protein or bathing solutions. Even if these theories deal with charge on the boundaries, they do so only in the particular case in which the far field voltage is clamped to zero.

These theories and simulations do not include the 'far fields' or general boundary conditions that produce resting or electrotonic potentials (Jack, Noble & Tsien, 1975) of nerve and muscle cells, or syncytical tissues (Eisenberg & Mathias, 1980) or long insulated conductors bathed in an ocean of salt water, for that matter (i.e., the Atlantic cable: Kelvin, 1855, 1856). They cannot describe the qualitative mechanism of the action potential (Hodgkin & Huxley, 1952; Hodgkin, 1971, 1992) because that involves feedback between the membrane potential (i.e., boundary potential) and the potential within and current flowing through a single channel molecule (Hodgkin, 1958). Much recent work on liquids, ionic solutions, and proteins has been devoted to dealing correctly with just the Dirichlet far field condition of (uniform) zero electrical potential at infinity (using Ewald sums or reactions fields, *viz.:* Friedman, 1975; Yoon & Lenhoff, 1990; Sharp, 1991; Zauhar, 1991; Gao & Xia, 1992; Saito, 1994; Tannor *et al.,* 1994; Rashen *et al.,* 1994; Chen *et al.,* 1994; Tironi, *et al.,* 1995). That work will soon be extended, no doubt, to include other boundary conditions of the electric field (Jackson, 1975; Griffiths, 1989), and of cells and tissues (Jack, Noble & Tsien, 1975; Eisenberg & Johnson, 1970; Mathias *et al.*, 1979) and of electrodiffusion, as it must, if it is to describe how macroscopic electrical, diffusional, or biological systems actually function.

***SOLVING THE PNP EQUATIONS MEANS DETERMINING THE ELECTRIC FIELD.*** The PNP system of equations can be solved simultaneously, without approximation, using the numerical methods described in the Appendix. Even though PNP is a quite complex, coupled set of nonlinear differential equations, the underlying physical principles are simple: all types of charge in the channel, and at its ends, have interacting effects on the distribution of potential and concentration in the channel. The shape of the electric field is found to change substantially, even qualitatively, by some $k_BT/e$ in most locations, in hundreds of different calculations simulating real experimental conditions, like those used in the laboratory, in which concentrations of ions are changed in the baths. Modulation of the electric field in channels is the rule, not the exception. As experimental conditions change, the potential profile changes.

Nonetheless, the surface of a channel or protein has often been described as a more or less unvarying set of potential barriers over which flux flows with a rate constant independent of the concentration of ions in the bath or channel (Hill, 1977; Walsh, 1979; Hill, 1985; Hille, 1992). Such models of flux over barriers (often described by rate constants) implicitly describe the surface of a protein as an unvarying potential profile—a 'potential of mean force' in the technical language of statistical mechanics—thereby ignoring the effects of shielding, even though those effects can be large at the ends of the channel (*loc. cit.* and see eq. (4)-(6) of the Appendix) and in its pore (*cf.* eq. (11) & (12) of the Appendix) and are large in many other physical systems (as described on. p. 9).

---

[18] or boundary conditions in general, following the philosophy stated clearly in Goldstein, 1980, p. 16: "On this [atomic] scale all objects, both in and out of the system, consist alike of molecules, atoms, or smaller particles, exerting definite forces, and the notion of constraint becomes artificial and rarely appears." This philosophy may be appropriate when atoms or clusters of atoms are isolated (either naturally or experimentally, so physicists can study their inherent properties without disturbance by the outside world). It clearly is ***in****appropriate in a system like a channel in a membrane which naturally interacts with the macroscopic electric field created by the *trans*-membrane potential.

**THE SURFACE OF A PROTEIN IS A DISTRIBUTION OF CHARGE, NOT POTENTIAL.** A channel (or a protein, for that matter) can certainly be described as a distribution of potential under one set of conditions. Indeed, it can be described as the same distribution of potential under another set of conditions, as long as all the charges (*cf.* p. 8) in the system stay the same. But if concentrations, or *trans*-membrane potential are changed (or if the protein binds a substrate or transmitter, or if it changes conformation), charges will change and so will the potential-of-mean-force.[19]

After any charge in the system changes, a profile of the potential-of-mean-force can still be used to describe the channel protein, but it will be a different profile with a different size and shape. The potential-of-mean-force might conceivably stay the same at one location, but only if that location is connected to a battery that supplies or withdraws charge to compensate for changes in other types of charge. The potential *profile* can stay the same only if many locations are connected to (different) batteries, each of which supplies or withdraws the charge necessary to keep its potential constant.

Describing a protein as a potential surface is equivalent to describing that surface as a source of potential, i.e., as a Dirichlet boundary condition, a surface where the potential is fixed. A surface of matter cannot be a source of potential, described by a Dirichlet boundary condition, unless it is a metal connected to a battery. The surface of a protein is neither metallic, nor c $\frac{5\mu\sec}{\text{Solvation time period} \doteq 10^{-14}} = 5 \times 10^8$ nnected to a battery and so the surface of a protein must be described as a distribution of permanent charge[20]. Only the permanent charge stays constant as experimental conditions vary and change the mobile charge (i.e., ions) in a channel and at its boundaries. If those mobile charges vary, the total charge varies, and so does the potential. Thus, as experimental conditions are varied, as the potential or concentrations in the bath are varied, the electric field varies because the charge in the channel's pore and at its boundaries vary, even if the (fixed) charge on the protein stays the same. The electric field is modulated by changes in experimental conditions: modulation is the rule, not the exception.

Because the fluxes through channels tend to be (combinations of) exponential functions of electrical potential[21], modulation of the electric field tends to govern the qualitative properties of channels. In this way, changes in the contents of the channel can govern the qualitative properties of current flow.

**CONFORMATION CHANGES AND THE ELECTRIC FIELD.** It is difficult to imagine a conformation change that does not depend on the potential profile in a channel protein. Thus, theories of conformation change—or traditional theories of open channel permeation, gating or mediated or active transport, for that matter—must recompute their rate constants (from the potential profile) as experimental conditions change and modify the electric field. Assuming rate constants independent of experimental conditions is likely to lead to serious error because rate constants tend to depend exponentially on potential (*cf.* eq. (11) & (12) of the Appendix). The underlying potential profiles depend strongly on experimental conditions (*cf.* eq. (9) of the Appendix) and this dependence is likely to be responsible for important biological functions.

---

[19] If the conformation changes, the distribution of permanent charge will change as well, of course.
[20] This is to a first approximation; to a second approximation, dielectric properties must also be included as they are in our analysis (*cf.* eq. 1).
[21] See eq. (10).

Rate and state models (Hill, 1977; Walsh, 1979; Hill, 1985) are both too flexible and too rigid, in my opinion. They are too flexible because the operational definition of 'state' (in a condensed phase dominated by friction, like a protein and its pore) is often vague and indeterminate: states and adjustable parameters can be added too easily to make a model fit too wide a range of data[22]. They are too rigid because the rate constants of barrier models are nearly always assumed to be independent of concentration of ions (or substrate) and that is nearly impossible, as we have mentioned before.

A protein or channel assumed to have an unchanging electric field is unrealistic because it almost certainly will violate the equations of the electric field[23]; it is also likely to be a dead protein, unable to respond to its environment, the way live proteins do. In my opinion, biological control mechanisms are likely to modulate the electric field in channels and proteins, thereby initiating and governing conformational changes, as well as driving 'catalysis', which is analogous to permeation through the open channel (Andersen & Koeppe, 1992; Eisenberg, 1990).

**VALIDITY OF PNP THEORY.**    Eisenberg, Klosek & Schuss, 1995 (following the simulations of Cooper, Jakobsson & Wolynes, 1985; Chiu & Jakobsson, 1989; and the lead of Barcilon, Chen, Eisenberg & Ratner, 1993) proved the validity of the Nernst-Planck (NP) part of the theory for discrete atomic systems without single filing. Barkai, Eisenberg, and Schuss, 1995, are extending the theory to single ion channels. Using only mathematics, Eisenberg, Klosek & Schuss, 1995, showed that the NP equations describe the mean of the probability density function for location in systems dominated by friction, in which atoms follow the Langevin model of atomic motion, i.e. Newton's laws plus fluctuations caused by atomic collisions.

The Langevin model is the starting place for most theories of atomic motion in condensed phases (Kramers, 1940; Gardiner, 1985). Even if the system is not entirely dominated by friction, and the Langevin equation is needed, or the friction has complex properties, and the generalized Langevin equation is needed (Hynes, 1985; 1986; Berne, Borkovec & Straub, 1988; Hänggi, Talkern & Borkovec, 1990; Fleming & Hänggi, 1993; Tuckerman & Berne, 1993), Eisenberg, Klosek & Schuss prove (using mathematics alone) that something very like Nernst-Planck is appropriate (their eq. 6.15). Surprisingly, the main results of their analysis depend only on mathematical identities. All that is necessary for their proof is the existence of conditional probabilities of location. The conditional probabilities can be directly derived from simulations of molecular dynamics (without using even the generalized Langevin equation, e.g., see Barcilon, Chen, Eisenberg & Ratner, 1993, Fig. 4 & 5) or from the Onsager-Machlup action formulation of Newton's laws in the presence of thermal agitation (Onsager & Machlup, 1953; see modern application: Elber, 1995).

---

[22] "…an excess of flexibility [in a theory] may well turn power into weakness. For a theory that explains too much ultimately explains very little. Its indiscriminate use invalidates its usefulness and it becomes empty discourse. Enthusiasts and popularizers, in particular, do not always recognize the subtle boundary that separates a heuristic theory from a sterile belief; a belief which, instead of defining the actual world, can describe all possible worlds." (Jacob, p. 22, 1982; *cf.* Monod, 1972, p. 69 *et seq.*). If Jacob and Monod had realized the relation of their work to that of Bardeen, Brittain, and Shockley; if they had known that rate constants can only be derived from nonlinear theories like $PB_n$ or PNP that have a wide range of "oriented, coherent, and constructive" behavior, perhaps the literature of allosteric models would be smaller. It presumably would not include so many papers that use underdetermined rate and state models to describe complex experimental systems of unknown structure.

[23] except under the conditions when constant field theory is a decent approximation (Chen and Eisenberg, 1991, and references therein).

Thus, there seems little question of the validity of the NP equations for a stochastic discrete atomic system. They describe the mean value of the probability density function of location of ions within the open channel. The question then is how to compute the mean potential used in the NP equations.

**THE MEAN POTENTIAL**. The Poisson equation can describe a potential on many time scales. In the PNP theory of channels, it describes the mean field approximation to the potential in the open channel. In particular, it is the potential averaged over the 5 μsec needed to measure a single digital sample of mean current (because of instrumentation and signal to noise problems). 5 μsec is much slower than the time scales important in solvation (0.01 psec, Stratt, 1995), or typical atomic motions in proteins (0.1 psec) and so averaging of these motions is necessary. 5 μsec is also slow compared to the time scale of ion permeation: a univalent ion crossing a membrane every 160 nsec carries 1 pA of current and a single ion takes roughly the same time to permeate a channel. (That is the mean first-passage time—MFPT—for ion permeation estimated by Cooper, Jakobsson & Wolynes, 1985; Chiu & Jakobsson, 1989; Barcilon, Chen, Eisenberg & Ratner, 1993; Eisenberg, Klosek & Schuss, 1995).

The time and length scale of atomic motions is known (Berry, Rice & Ross, 1980) from experiment (e.g., measurements of temperature factors of proteins by x-ray diffraction: Frauenfelder, Petsko & Tsernoglou, 1979; Parak and Knapp, 1984; Smith, *et al.,* 1990; Kuriyan *et al.,* 1991; Brooks, Karplus & Pettitt, 1988, McCammon & Harvey, 1987); from general theoretical considerations (e.g., kinetic theory, McQuarrie, 1976, Garrod, 1995); most vividly from the simulations of molecular dynamics in general (Burkert & Allinger, 1982; see Davidson, 1993, which introduces a review of the 'state of the art': Chem. Reviews, 93(7), 1993); from simulations of proteins (Brooks, Karplus & Pettitt, 1988, McCammon & Harvey, 1987); and from simulations of channel proteins, in particular gramicidin (reviewed in Roux & Karplus, 1994), including our own (Elber, Chen, Rojewska & Eisenberg, 1995). The range of time scales in proteins is remarkably large: experiments show conformation changes ranging from $10^{-15}$ sec to $10^4$ sec (Brooks, Karplus & Pettitt, 1988: Table I, p. 19). The interactions which determine the energetics of solvation seem to occur on a time scale of $10^{-14}$ sec (the large literature is reviewed in Stratt, 1995); $10^{-7}$ sec is a reasonable estimate of the MFPT of one ion permeating a channel (*loc. cit.*).

The RMS deviations of atoms from their mean positions are significantly larger (*op.cit.*), and the motions more violent, than imagined in traditional theories of channel permeation (Hille, 1992; Andersen & Koeppe, 1992; *cf.* Purcell, 1977; Berg, 1983). For example, comparing a snapshot of a simulation of a gramicidin channel with one taken some 10 psec later shows that most atoms are more than 1 Å away from where they were initially. If the electrostatic potential is computed from each snapshot, striking variations are found. The potential at most locations has changed more than $k_BT/e$: Coulomb's law shows that a 1 Å displacement of a carbonyl oxygen (containing ~0.6$e$ of charge), produces a potential change >1 $k_BT/e$ at distances of a few Ångstroms (e.g., Israelachvili, 1985, Ch. 3). Thus, potential profiles in pores change substantially at least every 10 psec; they are likely to fluctuate[24] (by several $k_BT/e$) many times indeed while one sample of open channel current is measured (~5 μsec).

It is the fluctuating potential that determines the current measured in channel experiments. The experimental measurement of current is taken over such a long time period (compared to atomic or ionic motions) that it is necessarily an average. Thus, the potential that determines the measured current must be an average too, of some sort or other. The functions being averaged vary wildly, containing some $\frac{5\mu\text{sec}}{\text{MFPT}\doteq 10^{-7}} = 50$ to $\frac{5\mu\text{sec}}{\text{Solvation time period}\doteq 10^{-14}} = 5\times 10^8$ fluctuations in the 5 μsec period of a single current measurement. The average cannot be a simple mean of the potential determined from individual trajectories because that average is numerically (inherently) ill-determined[25]. Rather, the average must be determined separately, *by its own theory*, just the way the average of the trajectories of a Brownian motion[26] are determined by Fokker-Planck equations, not by actually adding up the trajectories (Arnold, 1974; Schuss, 1980; Gardiner, 1985; Gard, 1988).

---

[24] These rapid fluctuations probably account for the very large open channel noise we (and others) have observed at high frequencies. Below (say) 10 kHz—where open channel noise has been reported—these rapid large fluctuations presumably 'average out' to zero.

The nature of the average potential depends on the recording system and atomic properties of the channel, as it does in any macroscopic condensed phase. If the motions of the permeating ion (or more precisely the permion) are much slower than the atomic motions surrounding it, the Hartree self-consistent field (SCF) approximation (Ashcroft & Mermin, 1976; Kittel, 1976) is usually invoked: the effective potential for a charged particle ion is determined from the averaged trajectories of the other charges, i.e. atoms.

It is certainly possible to imagine or build a system where macroscopic flux depends on some extremely weighted average or particular extreme value of the fluctuating atomic potential, and the SCF approximation fails. But if macroscopic properties depend on such extra-ordinary events, they cannot be described literally by classical macroscopic theories of average properties like Maxwell's equations and diffusion equations of Fick's law, or statistical mechanics as built on the Boltzmann transport equation. Classical macroscopic theories of matter are literally valid only when the averages of atomic trajectories are well behaved. Most theories of condensed phases, solid state, or gas phases assume that the mean dependent variables (like potential) obey the macroscopic laws of electrostatics, and the fluxes are described by macroscopic laws of diffusion of charged particles, i.e., the Nernst-Planck equations, even though the electrical potential and other parameters are known (by direct experimental measurement in many of these sciences) to fluctuate wildly on atomic scales of length and time. These theories[27] all use macroscopic laws like Poisson's equation and the NP equations to predict macroscopic measurements. They all use SCF and effective parameters to describe macroscopic properties as averages of atomic trajectories. And all these macroscopic theories fit a wide range of the phenomena seen in ordinary experiments—the classical phenomena of physics. Of course, they cannot describe the extraordinary phenomena of modern physics (e.g., conduction of current by holes in semiconductors), precisely because those phenomena depend on correlations that are ignored in the averages of traditional macroscopic theory.

---

[25] The slightest systematic error, or error in truncation of the sum, or round-off error would dominate the estimate of the average of such a function. The various averaging procedures of *equilibrium* statistical mechanics do not apply to these systems far from equilibrium, and attempts to use them have not been successful (Allen & Tildesley, 1990, Ch. 8-11; Haile, 1992: Ch. 8; Evans & Morriss, 1990; Hoover, 1991); indeed, such attempts have led some authors to propose a radical reworking of the 'thermodynamic' theory of flux (see experiments of Keizer & Chang, 1987, and Hjelmfelt & Ross, 1995; and discussion and references in Peng, *et al.*, 1995, and Vlad & Ross, 1994*a,b*; Keizer, 1987*a,b* ).

[26] which, mathematically speaking, are functions of unbounded variation, oscillating an infinite number of times in any finite interval, no matter how small (Wong & Hajek, 1985, p. 53) and thus are hard to evaluate in a finite simulation or calculation.

[27] *Boltzmann transport equation:* McQuarrie, 1976, Résibois and de Leener, 1977; Berry, Rice & Ross, 1980; Cercignani, 1988; Spohn, 1991; Balian, 1992; Cercignani Illner & Pulvirenti, 1994; Garrod, 1995; *condensed phase theory:* Kramers, 1940; Gardiner, 1985; Hynes, 1985; 1986; Hänggi, Talkern & Borkovec, 1990; Antosiewicz *et al.,* 1994; Sharp & Honig, 1990; Fleming & Hänggi, 1993; *solid state:* Ashcroft & Mermin, 1976; Cox, 1993); *gases:* Chapman & Cowling, 1970; Hirschfelder, Curtiss, and Bird, 1954; Mason & McDaniel, 1988.

**QUASI-PARTICLES AS CORRELATED MOTIONS IN A MEAN FIELD.** It is instructive to consider how such extraordinary phenomena are analyzed in other sciences (Ashcroft & Mermin, 1976; Mason & McDaniel, 1988; Cox, 1993) where direct measurements of atomic and transport properties are routine, along with extensive simulations (Bird, 1994). In these sciences, mean-field theories are common, despite the universality of correlated motions, and theories like PNP (or identical to PNP: Ashcroft & Mermin, 1976: Ch. 28 & 29) are used successfully to fit a wide range of qualitative and quantitative behavior (Mahan, 1993). The PNP equations appear because they arise from conservation laws (that are true quite no matter what are the details of the atomic motion) and simple constitutive laws, like Fick's law, which approximately describe a wide range of systems. But these mean field theories are often not literally true as descriptions of motions of individual atoms, because the atomic motions are extraordinarily correlated. Rather, the mean field theories are used[28] to describe the collective motions as if they arose from the rigid body translation of a group of atoms (Goldstein, 1980: Ch. 5). The (sometimes hypothetical) rigid body is called a 'quasi-particle'. One of the main objects in such fields is to establish the existence of such quasi-particles (by experiment, simulation, and theory) and to determine their conservation laws, laws of motion, and effective parameters.

Consider, for example, the extraordinary phenomena of ferromagnetism, superconductivity, and polarons (Ch. 33 & 34 of Ashcroft & Mermin, 1976; Mahan, 1993, Ch. 5 & 9; Ch. 6.2 of Cox, 1993). Classical mean field theories cannot describe these phenomena, but SCF is still used: correlated motions that were described improperly in the simplest mean field theories are analyzed in detail, and summarized as the motions of a quasi-particle in a SCF mean field. Examples of quasi-particles are not hard to find: a phonon, polaron, or Cooper pairs are such quasi-particles; indeed, even the hole and 'electron' of solid state physics are *not real particles*, but rather quasi-particles with mass and other properties quite different from real particles, for example, electrons in free space. The mean field approach is retained and extended to these quasi-particles, but the objects moving in the mean field are no longer the 'real' atoms or molecules themselves. Rather, they are groups of atoms whose correlated motions allow them to be described as quasi-particles with definite properties that follow their own law of motion, obeying the macroscopic laws of diffusion and electrostatics[29], albeit using effective parameters which are only indirect representations of the complex underlying atomic properties, and so have numerical values that are not immediately understandable (e.g., negative mass of an 'electron' in a semiconductor, Spenke, 1958, p. 58-60; Kittel, 1976, p. 220).

It is actually *necessary* to construct a theory of the correlated motions, instead of the atomic motions in many cases, because the numerical averaging needed to link atomic motions and experimental observations is impossible to actually perform. Indeed, in some cases (when flux flows or dimensionality is reduced: Allen & Tildesley, 1990, Ch. 8-11; Haile, 1992: Ch. 8; Evans

---

[28] A mathematician might say "they are used figuratively" when speaking of those cases where the quasi-particle and its behavior cannot be rigorously derived from the underlying atomic dynamics.

[29] The physicist's definition of a quasi-particle is analogous to the biochemist's definition of a (conformational) state of a protein. The quasi-particle follows Langevin equations (Newton's laws of motion with noise added), although sometimes with peculiar conservation laws; the chemical state follows the law of mass action; in both cases the existence of the correlated motion (quasi-particle or state) is assumed. The law of mass action (in condensed phases) can only be derived (as illustrated in eq. 11 & 12 of Appendix) from Langevin equations (*loc. cit.*); and so the physicist's approach has significant advantages, at least in my view.

& Morriss, 1990; Hoover, 1991; Lowe, Frenkel & Masters, 1995), the average may not converge to a definite value no matter how long the averaging goes on, and so the average may not exist, in the mathematical sense of the word. Consider the current through an open channel that is measured in 5 μsec. It is numerically impossible to actually calculate the average of the rapidly fluctuating (estimate of its) correlation function over the 50 to 500 million fluctuations that occur during that measurement, because round-off error would inevitably dominate such an average. Rather, a theory of the average may have to be derived analytically to replace the uncomputable average of the trajectories or correlation functions.

**QUASI-PARTICLES AND PNP.** It is well to remember this experience of other sciences as we try to apply PNP to channels. We too are likely to find a range of phenomena (e.g., of the open channel) that are well described by the crude mean field theory presented in the Appendix, but with effective parameters whose meaning can be difficult to sort out, because the objects described by the mean field theory are not atoms but quasi-particles we call 'a permion' (Elber et al., 1995), that move along an effective reaction coordinate, a twisted path, that is usually not perpendicular to the membrane surface. Indeed, some of the classical phenomena of open channel permeation arise from the correlated motion of the permeating ion and water as transport experiments showed some time ago: Rosenberg & Finkelstein, 1978a,b; Levitt et al., 1978; Finkelstein & Andersen, 1981; Dani & Levitt, 1981; Levitt, 1984; Finkelstein, 1987; Hille, 1992. The correlated motions are, of course, also apparent in simulations (Chiu et al., 1989; Roux & Karplus, 1991; Roux & Karplus, 1994; Elber, Chen, Rojewska & Eisenberg, 1995). Other classical phenomena depend on the interactions of ions coming from different sides of the membrane (e.g., non-ideal ratios of unidirectional fluxes of tracers, Jacquez, 1985; Hille, 1989; Chen and Eisenberg, 1993b) and so must be described by a theory that allows such interactions, for example, a mean field theory of the correlated diffusion of a permion in a single file (Barkai, Eisenberg & Schuss, 1995). Of course, the utility of this idea of a permion remains to be established. Now, it is a compelling and not very new image; eventually, it may become a full fledged theory predicting flux, fluctuations, and selectivity.

It is even possible that the phenomena we call (single channel) gating will be best described as the motion of a quasi-particle (perhaps, the permion; more likely a quasi-particle with different properties, a 'gation') that follows its own laws of motion (i.e., Langevin equation) with probability density described by something like PNP. If that motion occurs over a high barrier, its mean time course will be exponential and rate theory will be a good approximation, as it is to most phenomena of single channel gating (McManus & Magleby, 1991; McManus & Magleby, 1988; McManus, *et al.*, 1988).

PNP is thus an appropriate mean field theory of the rapidly fluctuating atomic scale potentials of the open channel, as long as averages are taken on the time scale at which current is measured (e.g., over 5 μsec, which is long compared to the permeation time of a single ion), and as long as the system is reasonably homogenous for that 5 μsec.

**HOMOGENEITY OF THE OPEN STATE.** The homogeneity of states of proteins cannot be assumed (Frauenfelder 1985, Ansari *et al.*, 1985; Frauenfelder, Sligar & Wolynes, 1991) and so it is fortunate that the homogeneity of the open state of a channel is known directly from experimental measurements of single channel currents. Measurements of open channel noise (chiefly from Sigworth's lab, starting with Sigworth, 1985; for more recent references see Heinemann & Sigworth, 1991, also Hainsworth, Levis & Eisenberg, 1995) show no correlation between successive samples of the current records: currents hardly vary within one channel opening or from opening to opening, or even experiment to experiment (when recording conditions are the same). The va-riance and power spectrum within a single prolonged opening is the same as the variance from opening to opening (*op.cit.*).

The homogeneity of currents observed experimentally implies an underlying structural homogeneity of the open state of channel proteins. The conformation and the shape of the electric field must be reasonably constant and nearly the same whenever the channel is open; otherwise, the experimentally measured currents would not be reasonably constant and nearly the same.[30]

The question is what do the words 'reasonable' and 'nearly the same' mean? I believe this question can be answered directly from experimental data because the open channel noise itself is a measure of the inhomogeneity of the open state. Indeed, it is an upper bound on the inhomogeneity of the open state because open channel noise can arise in other ways besides inhomogeneity.

In fact, the measured current is a more sensitive measure of the homogeneity of the open state than the potential profile: current through a channel is an exponential function of potential in nearly all theories (e.g., as shown explicitly in eq. (10), left hand side, Appendix), and the concentration is also a steep function of potential as shown on the right hand side of eq. (9); consider special case with $C_j(R) = 0$, $V_{appl} = 0$. Thus, if single channel current is found to be reasonably constant experimentally, the contents of the channel are likely to be reasonably constant, and the potential profile is likely to be even more constant.

The homogeneity of the open state is particularly interesting, because states of proteins in general are not considered so homogeneous (*op.cit.*). The homogeneity of the open state is probably a consequence of the high density of fixed (i.e., permanent) charge along its wall. If that charge is referred to the volume of the channel's pore, its concentration is several molar; e.g., in gramicidin, in which the carbonyl oxygen's make up the wall of the channel, each 3Å turn of the helix contains about 0.6*e,* giving a concentration of permanent charge of 2.6 Molar. Such a highly charged tunnel is likely to change potential dramatically if its contents change and leave substantial fixed charge unshielded. It could well be described as the ionic wire postulated by Rosenbusch (1988).

We should thus not be surprised that open channel noise is reasonably well behaved in most channels. It does not differ from instrumentation noise by more than a factor of 2× in the papers of Sigworth (where it is often less than that) or Hainsworth, or in most other channels. Indeed, if uncertainty in the level of open channel current is comparable to the mean amplitude of open channel current, or if the current fluctuates significantly (say RMS deviation > 10% of open channel current) on a time scale comparable to the (mean) duration of open channel current, the records do not fit within the paradigm of single channel recording (Bean, *et al.,* 1969; Hladky & Haydon, 1970; Neher & Sakmann, 1976; Ehrenstein & Lecar, 1977; Sigworth & Neher, 1980; Neher, 1982; Sakmann & Neher, 1983, 1995) and are likely neither to have been pursued (very far) nor to have been reported (in full length publications).

***PNP AS A MEAN FIELD THEORY.*** We conclude that the potential of PNP theory is a well-defined mean potential appropriate to describe current flowing in open channels on time scales longer than some 5 μsec. PNP should be viewed as the first order SCF theory of flux in channels. It does for channels, in the presence of flux, what $PB_n$ theory has done so successfully for proteins in the absence of flux (for some time now). PNP is a practical theory because of the advances in nu-merical analysis that allow quick computation of the full coupled nonlinear system. As far as we know, PNP is the first channel theory to solve the full set of equations involving induced, surface, and fixed charge, and flux through the channel.

---

[30] The subconductance states and flickers commonly seen in wide bandwidth recordings of channels also set bounds on the validity of our argument: they should be viewed provisionally as a measure of another kind of ***in***homogeneity of the open state, distinct from that observed in measurements of open channel noise.

***GENERAL RULES:*** our analysis implies general rules independent of the limitations of mean-field, or quasi-particle theories. I believe *any* theory must obey these rules, if it is to be consistent with the properties of the electric field.

(1) The theory must compute the electric field, not assume it.
(2) The theory must compute the electric field from all types of charge (*cf.* p. 8).
(3) The theory must never assume that a potential is maintained constant at some location in matter (as experimental conditions are changed), unless that location is connected by a wire to experimental apparatus which serves as a source of energy and charge (i.e., a battery or amplifier).

Anything that changes the electric field is almost certain to change rate constants; thus,

(4) The theory must recompute rate constants when experimental conditions change, remembering that a protein or channel with an unchanging electric field is likely to be a dead protein, unresponsive to its environment, at least compared to natural proteins or channels.

***CHECKING TRADITIONAL THEORIES.*** A traditional theory can easily be checked to see if it is consistent with the laws of electricity: it is consistent if and only if the potential profile satisfies Poisson's equation (and the rate constants in it are computed from the potential profile, (*cf.* eq. (11) & (12) of the Appendix). In one-dimension this means the second spatial derivative (the curvature) of the potential must equal[31] (at every location) the sum (at that location) of all charges in the model, including partial charges found on nearly every atom of a protein (see p. 6 & 20). If the charges in the model do not add up to the second derivative, the model is inconsistent and incorrect. This check has to be repeated in each experimental situation (e.g., for each set of bath concentrations and potentials and at each location in the system).

A traditional theory can easily be checked to see if it is consistent with our ideas of condensed phases (see footnote 3). Flux in a condensed phase always depends on friction; it is wise to check that the theory being used displays the dependence of flux on the diffusion constant or a derived parameter. Many do not (Hill, 1977; Walsh, 1979; Hill, 1985; Hille, 1992).

***PROFILE OF PERMANENT CHARGE.*** The profile of permanent charge is perhaps the most interesting effective parameter of PNP theory, because it determines the qualitative properties of the open channel (as doping determines the qualitative properties of transistors, along with the bias voltages) and because it can be modified so easily nowadays by changes in the amino acid sequence of proteins once the gene for that protein is known and cloned. Of course, only the primary structure—the sequence of amino acids—can be read from the genome: although that sequence is thought to determine the three dimensional folding pattern of the polypeptide chain, only nature knows how to do that (Creighton, 1992): no one can predict three dimensional structure (indeed even refine a bad guess) at the present time. We have seen (on page 16) how sensitive the potential profile is to the location of charge, and so we can see how hopeless it is to guess the biologically relevant profile of permanent charge without knowing the three dimen-sional structure, heroic attempts notwithstanding. Nonetheless, *changes* in the primary sequence, which may be assumed (with various degrees of certainty) to leave the rest of the three dimensional structure essentially unchanged, can often be interpreted. Indeed, some changes have strikingly specific results, presumably because they mimic a single amino acid substitution that evolution found useful, sometime ago.

---

[31] ignoring units, for the sake of simplicity in writing.

Successful and complete investigation of structure function relations, of course, will need measurement of structure. Three-dimensional structures are much harder to resolve than primary sequences, but some channel structure are now known at atomic resolution (e.g., Cowan, *et al.,* 1992) and more will be. The question then is how do we determine the one dimensional distribution of permanent charge of PNP theory from the three dimensional structure of the channel protein (on the one hand) or the current voltage relations of the open channel (on the other)? As significant as these questions are, I do not know the answers (or even if answers exist, in any general sense). It is best now simply to ask these questions, waiting for another day, when perhaps answers will be available, at least for a specific channel.

**POSTSCRIPT.**   I end on a personal note. As a physicist, I distrust theories that constrain a protein to one potential profile, because they are incompatible with the fundamental properties of the electric field. As a biologist I distrust theories that cannot produce the repertoire of behaviors needed if proteins are to produce the "oriented, coherent, and constructive activity" of the vital molecules of life (Monod, 1972: p. 45). As a channologist, I know that conformation changes occur and are biologically important: indeed, that is why I believe they are worth describing and analyzing. Conformation changes, of course, arise from physical forces, despite their vital functions.

Perhaps the variation of the electric field with conditions—which (along with doping) allows a lump of silicon to be a computer—allows channels, enzymes, and proteins to perform their biological functions. Perhaps, the change of shape of the field produces some of the vital behavior that conformation changes were invented to describe. Only computing the field can tell us if this is so.

# APPENDIX: PNP THEORY

The historical antecedents to PNP and its relationship to $PB_n$ theory (e.g., of shielding), which is close, have been discussed in the text (p. 9). Our contributions have been reported in a series of papers starting with Barcilon, 1992. Barcilon considers the full three dimensional problem and derives the appropriate form of the perturbation expansion. Incidentally, (in a result overlooked by many biophysicists) he solves analytically the electrostatic problem of a finite length cylinder embedded in a thin membrane, giving two different exact expressions for the potential. Barcilon *et al.*, 1992, and Chen *et al.*, 1991, derive and solve the one-dimensional theory without permanent charge. They show when the approximation of constant (electric) field and constant gradient (of concentration) can approximate the full equations (without permanent charge). Chen and Eisenberg, 1993*a*, put permanent charge in the theory; Chen and Eisenberg, 1993*b*, introduce nonequilibrium boundary conditions that allow a channel of one structure to produce single filing, and flux coupling reminiscent (in some ways) of mediated transporters. Interestingly, it has just come to our attention that the theory of bulk ternary ionic solutions includes flux coupling (Wendt, 1965) *of positive or negative sign* (Vitagliano & Sartorio, 1970). Eisenberg, 1995, embeds PNP theory in a hierarchy of models of different resolutions; Park, Barcilon, Chen, Eisenberg & Jerome, 1995, analyze the qualitative properties of the theory in the absence of permanent charge and built-in potentials. Chen, Eisenberg, Jerome & Shu (1995), generalize PNP theory to allow changes in temperature and pressure and the resulting transport of heat and mass. Eisenberg, Klosek & Schuss (1995) derive (just) the Nernst-Planck equations from a stochastic analysis of flux over barriers begun in Barcilon, Chen, Eisenberg & Ratner, 1993. Barkai, Eisenberg & Schuss, 1995, extend the stochastic analysis to a one ion channel.

Because of this extensive documentation, here we only state the main equations used to predict the results of typical experiments. In a typical experiment, the current *I* through a single open channel is studied as a function of the trans-membrane potential applied to the baths $V_{appl}$ in a variety of solutions of different composition $C_j(0)$ and $C_j(d)$. The theory used to predict this current starts with **Poisson's equation** (written here in dimensional form), which determines the potential $\varphi$ (units: volts) from the charges present (*cf.* p. 8)

$$-\varepsilon_{H_2O}\varepsilon_0 \frac{d^2\varphi}{dx^2} = \overbrace{eP(x)}^{\text{Permanent Charge}} + \overbrace{e\sum_j z_j C_j(x)}^{\text{Channel Contents}} + \overbrace{\underbrace{\tilde{\varepsilon}\left[\Delta(1-x/d)-\varphi(x)\right]}_{\text{Deviation from Constant Field}}}^{\text{Induced Charge}} \tag{3}$$

The concentrations $P(x)$ and $C_j(x)$ are numbers of particles per unit volume, e.g., cm$^{-3}$. The dielectric properties of the channel protein and its watery pore (radius *r*, length *d*) are described by the permittivity of free space $\varepsilon_0$ (units: coulombs$\cdot$volt$^{-1}\cdot$cm$^{-3}$); the (dimensionless) dielectric constants $\varepsilon_p$ and $\varepsilon_{H_2O}$, respectively, and the effective dielectric parameter $\tilde{\varepsilon} \equiv \frac{2\varepsilon_p \varepsilon_0}{r^2 \ln(d/r)}$, defined and derived in Barcilon, 1992; discussed Chen & Eisenberg, 1993*a*.

***DONNAN POTENTIAL.*** Permanent charge at the ends of the channels creates Donnan or built-in potentials *in the baths* $\Phi_{bi}(0)$, $\Phi_{bi}(d)$ (dimensionless: $\Phi \equiv e\varphi/k_B T$. These are the surface potentials stu-died extensively in membrane biology (McLaughlin, 1989; Green and Andersen, 1991) and are easily computed because the bathing solutions are made of (nearly) equal amounts of cations and anions, *viz.*, $\sum_j z_j C_j(L) = \sum_j z_j C_j(R) = 0$. Then,

$$\Phi_{bi}(0) \equiv \log_e \frac{\sqrt{P^2(0) + 4C_1(L)C_2(L)} + P(0)}{2C_2(L)} ; \tag{4}$$

$$\Phi_{bi}(d) \equiv \log_e \frac{\sqrt{P^2(d) + 4C_1(R)C_2(R)} + P(d)}{2C_2(R)} \tag{5}$$

The potentials on each end of the pore, and from one end to the other, are

$$\begin{aligned}\Phi(0) &= \Phi_{bi}(0) + V_{appl} \\ \Phi(d) &= \Phi_{bi}(d) \\ \Delta &\equiv \varphi_{bi}(0) - \varphi_{bi}(d) + \tfrac{k_B T}{e} \cdot V_{appl}\end{aligned} \tag{6}$$

Note that the potential $\Delta$ is *not* the *trans*-membrane potential $V_{appl}$ applied to the baths. The baths are assumed at equilibrium, even when current flows, so

$$\begin{aligned}C_j(0) &= C_j(L)\exp\left[-z_j \Phi_{bi}(0)\right] \\ C_j(d) &= C_j(R)\exp\left[-z_j \Phi_{bi}(d)\right]\end{aligned} \tag{7}$$

**Nernst-Planck equations** determine the flux $J_j$ of each ion

$$J_j = -D_j \left[ \overbrace{\frac{dC_j}{dx}}^{\text{Diffusion}} + \overbrace{z_j C_j(x) \frac{d\Phi}{dx}}^{\text{Migration}} \right] \tag{8}$$

where $D_j$ is its diffusion constant.

The Nernst-Planck equations (7) can be integrated analytically to give expressions for the concentration of ions in the channel, namely the channel's contents

$$C_j(x) = \frac{C_j(L) \cdot \exp z_j [V_{appl} - \Phi(x)] \cdot \int_x^d \exp z_j \Phi(\zeta) d\zeta}{\int_0^d \exp z_j \Phi(\zeta) d\zeta} \\ + \frac{C_j(R) \cdot \exp[-z_j \Phi(x)] \cdot \int_0^x \exp z_j \Phi(\zeta) d\zeta}{\int_0^d \exp z_j \Phi(\zeta) d\zeta} \tag{9}$$

This system of equations (5) & (9) must be solved simultaneously because the potential depends on the concentrations $C_j(x)$ through the Poisson equation (5) but the concentrations also depend on the potential through the integrated Nernst-Planck equations (9). Indeed, the concentrations depend exponentially on the potential. Once the potential profile is set, the distribution of con-centration is determined. In that sense, assuming a potential profile is equivalent to locking a channel into a specific occupancy state, from which it cannot move as long as the potential profile does not change.

A different integration of the Nernst-Planck equations shows that flux (and the observable, the electrical current $I$) also depends exponentially on the potential profile.

$$J_j = D_j \frac{C_j(L)\exp(z_j V_{appl})}{\int_0^d \exp z_j \Phi(\zeta) d\zeta} - D_j \frac{C_j(R)}{\int_0^d \exp z_j \Phi(\zeta) d\zeta} ; \qquad I = \pi r^2 \cdot \sum_j e z_j J_j \qquad (10)$$

While the numerators of these terms can be written as functions of (just) the electrochemical potential, the denominators cannot (Chen and Eisenberg, 1993*b*; Eisenberg, 1995). The denominator and flux itself depends exponentially on $\Phi(\zeta)$, the electrical potential profile itself (not the electrochemical potential), which in turn depends on all the variables and parameters of the system through Poisson's equation (1). Once the potential profile is set, the flux is determined. In that sense, assuming a potential profile is equivalent to locking a channel into a specific conducting state, from which it cannot move as long as the potential profile does not change.

Each term of equation (10) describes the (so-called) unidirectional flux measured by tracers (usually radioactive isotopes) moving into a medium of (nearly) zero tracer concentration (see precise definition of unidirectional flux in Chen and Eisenberg, 1993*b*, and reference to the extensive literature in, for example, Jacquez, 1985). The fluxes can also be written as a chemical reaction (without approximation, for any potential barrier provided concentration boundary conditions are in force)

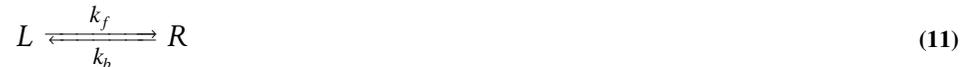

$$L \underset{k_b}{\overset{k_f}{\rightleftarrows}} R \qquad (11)$$

in which the rate constants are the conditional probabilities of the underlying diffusion process described by a full Langevin equation and boundary conditions (Eisenberg, Klosek & Schuss, 1995).

$$k_f \equiv \text{Prob}\{R|L\} = \frac{D_j \exp(z_j V_{appl})}{\int_0^d \exp z_j \Phi(\zeta) d\zeta} ; \quad k_b \equiv \text{Prob}\{L|R\} = \frac{D_j}{\int_0^d \exp z_j \Phi(\zeta) d\zeta} \qquad (12)$$

These expressions are not as obvious as they seem[32]: the *precise* definition of the conditional probabilities and *their specification* in terms of the full Fokker-Planck equation and boundary conditions took us some eight years (Barcilon, Chen, Eisenberg & Ratner, 1993; Eisenberg, 1995), only at the end of which was the simple expression (11) derived[33].

**NUMERICAL SOLUTION.** The design and choice of the numerical procedure is critical for the success of the calculations (Fatemi, Jerome & Osher, 1991: Jerome, 1995). The procedure was developed independently by Chen, e.g., Chen & Eisenberg, 1993*a*. Severe numerical difficulties arise with other methods of solving the system of eq. (3) & (8) and boundary conditions (4)−(7) (e.g., successive approximation to integral equations: Chen, Barcilon & Eisenberg, 1991; stiff differential equation solvers: see electrochemistry literature, chiefly from Buck and Mafé's group, *op.cit.*).

---

[32] Indeed, it is still not known whether diffusion can be written as a chemical reaction (without approximation for any potential profile) for any other boundary conditions.

[33] Eq. (12) surprised me, and amused many of my collaborators, given my oft stated opinions, if not prejudices, about rate and state models (*e.g.*, Cooper, Gates & Eisenberg, 1988*a,b*).

The coupled nonlinear system of equations (5) & (9) is solved numerically by substituting the integrated Nernst-Planck equation (9) into a discretized version of the Poisson equation (5). We start with a reasonable initial guess, say (a discretized version of) the constant field potential

$$\varphi(x; \text{initial guess}) = \left(1 - \tfrac{x}{d}\right)\left[\tfrac{k_B T}{e} V_{appl} + \varphi_{bi}(0) - \varphi_{bi}(d)\right] + \varphi_{bi}(d) \tag{13}$$

That initial guess of the potential profile is substituted into the right hand side of (a discretized version of) the integrated Nernst-Planck equation (9) to determine the congruent initial guess of concentration $C_j(x; \text{initial guess})$ and that guess is substituted into the right hand side of Poisson's equation, which is then solved (it is linear!). The resulting estimate of the potential $\varphi(x; \text{first iterate})$ is substituted into the integrated Nernst-Planck equation (9) and so determines a first-iterate of the concentration profiles $C_j(x; \text{first iterate})$. These two first-iterates $\varphi(x; \text{first iterate})$ and $C_j(x; \text{first iterate})$ are substituted into the right hand side of Poisson's equation (3), which is again solved, now to determine the second-iterate $\varphi(x; \text{second iterate})$, a better approximation to the potential profile. The second-iterate of potential determines a second-iterate of concentration by equation (9); together, the two second-iterates determine the third-iterate of potential, and so on for ten iterations which typically converge in less than one second to better than one part in $10^{12}$ on our workstation, an IBM RS/6000 Model 550. Each iteration involves a solution of the (discretized version) of Poisson's equation at 4,000 points of a uniform spatial mesh and yields a potential profile $\varphi(x)$, and concentration profiles $C_j(x)$ for one applied potential $V_{appl}$ and one set of concentrations $C_j(L)$ and $C_j(R)$. From these profiles, the flux $J_j$ and current $I$ at that potential $V_{appl}$ and for those concentrations are calculated by equation (10). Repeating the calculation at different potentials produces an IV curve in some 100 seconds of computer time if $I(V_{appl}; C_j[L]; C_j[R])$ is determined at 100 values of $V_{appl}$. Of course, a different IV curve is computed for each set of concentrations $C_j(L), C_j(R)$.

## ACKNOWLEDGMENT

This paper describes the work of my collaborators as much as my own: it is a joy to have shared this adventure with Duan Chen and our extraordinary co-workers, Victor Barcilon, Ron Elber, Joe Jerome, Mark Ratner, and Zeev Schuss. It is an even greater joy to thank them. My colleagues at Rush have contributed in more ways than I can enumerate: I am particularly grateful to Eduardo Rios for helping me interact with the classical traditions of the field and for his many other questions and suggestions (which stimulated, among other things, footnotes 11, 13, 25).

More specific thanks are due to Lou DeFelice, who suggested that I write this paper; to Joe Blum, who requested the Appendix; to Duan Chen, who suggested the metaphor of the dead channel locked into one occupancy state; to Olaf Andersen, who motivated me to relate PNP to traditional treatments of surface charge; to Tom DeCoursey, who suggested the ideas in footnote 10 and 30, amongst others; and to Richard Henderson, who forced me to think clearly about fluctuations, their means and meaning; and to Eberhard von Kitzing for sending me a manuscript before publication.

I have no words that can express my gratitude for the steadfast support of Dr. Andrew Thomson and the National Science Foundation. Without that, this work could not have been done.